\documentclass[12pt]{article}
\pdfoutput=1
\usepackage{hyperref}
\usepackage{cite}
\usepackage{color}
\usepackage{graphicx}
\usepackage{subcaption}
\usepackage{amsmath}
\usepackage{amssymb}
\usepackage{xspace}
\usepackage{verbatim}
\usepackage{slashed}
\usepackage{mathtools}

\usepackage[normalem]{ulem}

\makeatletter
\@addtoreset{equation}{section}

\makeatletter
\renewcommand\section{\@startsection {section}{1}{\z@}%
                                   {-3.5ex \@plus -1ex \@minus -.2ex}
                                   {2.3ex \@plus.2ex}%
                                   {\normalfont\large\bfseries}}
\renewcommand\subsection{\@startsection{subsection}{2}{\z@}%
                                     {-3.25ex\@plus -1ex \@minus -.2ex}%
                                     {1.5ex \@plus .2ex}%
                                     {\normalfont\bfseries}}

\def\baselinestretch{1.2}
\newcommand{\be}{\begin{equation}}
\newcommand{\ee}{\end{equation}}
\newcommand{\beq}{\begin{eqnarray}}
\newcommand{\eeq}{\end{eqnarray}}

\newcommand{\gone}[1]{{}}

\def\os{\overline{s}}

\begin{document}
\begin{titlepage}
\begin{flushright}
MAD-TH-16-07
\end{flushright}

\vfil

\begin{center}

{\bf \Large
Weak Gravity Conjecture and Extremal Black Holes
}

\vfil

William Cottrell$^{1}$, Gary Shiu$^{1}$, Pablo Soler$^{2}$

\vfil

${}^{1}$ Department of Physics, University of Wisconsin, Madison, WI 53706, USA\\
 ${}^{2}$ Institut f\"ur Theoretische Physik, Universit\"at Heidelberg,\\
 Philosophenweg 19, D-69120 Heidelberg, Germany

\vfil

\end{center}

\begin{abstract}
\noindent Motivated by the desire to improve our understanding of the Weak Gravity Conjecture, we compute the one-loop correction of charged particles to the geometry and entropy of extremal black holes in 4d. We find that fermion loops provide evidence for the necessity of the `magnetic' WGC cutoff.  Moreover, for a certain regime of black holes, we find entropy corrections with unusual area scaling. The corrections are reduced when supersymmetry is present, and disappear in ${\cal N}=4$ supergravity. We further provide some speculative arguments that in a theory with only sub-extremal particles, classical Reissner-Nordstrom black holes actually possess an infinite microcanonical entropy, though only a finite amount is visible to an external observer.   

\end{abstract}
\vspace{0.5in}

\end{titlepage}
\renewcommand{\baselinestretch}{1.05}  

\section{Introduction}

The Weak Gravity Conjecture (WGC) \cite{ArkaniHamed:2006dz} is a tantalizing statement whose implications in particle physics and cosmology have been actively discussed in recent years.  In essence, the conjecture states that any quantum mechanically consistent Einstein+Maxwell theory should have superextremal particles.  This may be generalized to theories with multiple $U(1)$ gauge fields \cite{Cheung:2014vva} as well as other p-form fields in various dimensions \cite{ArkaniHamed:2006dz, Brown:2015iha, Heidenreich:2015nta} and to holographic setups~\cite{Nakayama:2015hga}.  The utility of the conjecture lies in the fact that it provides a simple criteria for when certain low energy models should be banished to the `swampland' \cite{Vafa:2005ui}, even if there is no clearly visible field theoretical inconsistency.  One can then use this criteria to restrict attention to models which potentially have a consistent UV embedding.

As has been extensively discussed \cite{ArkaniHamed:2006dz, delaFuente:2014aca, Rudelius:2014wla, Rudelius:2015xta, Montero:2015ofa, Brown:2015iha, Bachlechner:2015qja, Hebecker:2015rya, Brown:2015lia, Junghans:2015hba, Heidenreich:2015wga, Kooner:2015rza,Conlon:2016aea,Hebecker:2016dsw, Saraswat:2016eaz}, the WGC potentially provides stringent constraints on axion inflationary models. These constraints, which apply to models with multiple axions, take the form of a correlated bound on the axion field range and the instanton action that generates the axion potential. Thus, the WGC casts doubts on whether corrections to the axion potential are safely suppressed for many axion inflationary models whose potential is generated solely by non-perturbative effects.
The conjecture may also provide some constraints on monodromy models \cite{Brown:2016nqt, Ibanez:2015fcv, Hebecker:2015zss,Herraez:2016dxn} (see also~\cite{Palti:2015xra, Andriot:2015aza, Baume:2016psm,Klaewer:2016kiy,McAllister:2016vzi,Valenzuela:2016yny} for related ideas). More recently, the WGC has also been applied to the study of (non-supersymmetric) vacuum stability~\cite{Ooguri:2016pdq,Freivogel:2016qwc,Danielsson:2016mtx}.

Unfortunately, not so much progress has been made in understanding the deeper reasons behind the WGC (see nevertheless~\cite{Cheung:2014ega, Harlow:2015lma, Bellazzini:2015cra, Heidenreich:2016aqi, Montero:2016tif}).  In fact, many of the arguments provided in the literature are either misleading, incomplete or incorrect.  For instance, one of the original motivations for the Weak Gravity Conjecture \cite{ArkaniHamed:2006dz} was to avoid `remnants' \cite{Susskind:1995da} and the closely related `species problem' \cite{Bekenstein:2002cha, Bousso:2004kp}. The reasoning is essentially as follows:  Imagine that we live in a world where WGC is violated and all charged particles are subextremal, i.e.,  $m > \sqrt{2} q M_{P}$.  In this case, any charged black hole will decay to an extremal Reisnner-Nordstrom solution, but no further.  Now, suppose additionally that $e \ll 1$, say, $e \sim 10^{-100}$.  Now, the argument goes, there would be $\mathcal{O}(10^{100})$ different extremal black holes in the mass range $0< M < M_{P}$ and these would give rise to problems analogous to those encountered in remnant theories.  For instance, it has been claimed \cite{Banks:2006mm} that the presence of so many states would entail a violation of the covariant entropy bound since many of these states with slightly different charges should be considered as macroscopically indistinguishable.

It is difficult to see how this argument could be made rigorous.  In fact, there is no known lower bound on the precision to which charge can be measured \footnote{One may, for instance, perform a simple electron scattering experiment and measure the deflection angle at low energy with arbitrary precision.}.  Thus, there is no a-priori reason to lump different charged states into the same ensemble.   Secondly, this reasoning would at best imply a bound up to some fuzzy order one quantity while string-theoretic examples suggest a sharp version of the conjecture determined precisely by the extremality bound. 

A related argument in support of the WGC is based on the so-called `species problem'. This refers to the possibility that by having a very large number of species we could violate the covariant entropy bound at finite temperature  \footnote{To see this, simply apply the Sackur-Tetrode equation to an ideal gas comprised of many species and observe that the covariant entropy bound may apparently be violated at low density.}.   It should be noted, however, that the case for a species problem has been somewhat undermined over time \cite{Casini:2008cr, Larsen:1995ax, Cooperman:2013iqr}.  In fact, it is understood that a large number of species can in fact be accomadated by a renormalization of Newton's constant.  Thus, as $S$ increases with the number of species, $G_{N}$ will decrease in such a way as to compensate, generically ensuring $S \le A/4G_{N}$.   Although string theory seems to prefer a small number of species, we are unaware of any a-priori evidence that this is necessary.

It is important to keep in mind that the species problem can only be evaded as described above if the number of species in a given mass range is {\it{finite}}. In contrast, the `remnant problem' \cite{Susskind:1995da} is more serious in that it entails an {\it{infinite}} number of remnants in a finite mass range.  For example, a theory with relatively irrational charges such as the one described in  \cite{Banks:2010zn} would imply a remnant problem.  In contrast,  references \cite{ArkaniHamed:2006dz, Banks:2006mm} merely imply a finite number of states in a given mass range and so it is not clear if the anti-remnant arguments of \cite{Susskind:1995da} are applicable at all.  Finally, the arguments as presented above require taking both $M \ge \sqrt{2} e M_{P}$ {\textit{and}} $e$ to be very small, while the conjecture is supposed to be valid even for $e$ of order, say, $0.1$.  

There thus seems to be some serious gaps in the present understanding for why the conjecture ought to be true.  To date, the best motivation for the WGC is purely empirical; all known compactifications of string theory satisfy some form of the conjecture.  

In this note we hope to improve upon the situation somewhat by outlining what we believe to be a more coherent, though still speculative, argument for the conjecture.  More precisely, we point out that simple classical arguments suggest that theories which violate the WGC conjecture likely suffer from a remnant problem, and not merely a species problem.  As a consequence, such theories may well violate entropy bounds such as \cite{Bekenstein2005, Bousso:1999xy} and the (strong) cosmic censorship conjecture \cite{Dafermos2005}.  In fact, these are likely related problems since a naked singularity can store a finite entropy with no horizon.  Moreover, when either an entropy bound or cosmic censorship fails we are immediately faced with other problems, such as unsuppressed black hole pair production or a failure of unitarity.  

These statements would seem to be in tension with various proofs of the covariant entropy bound, though we remind the reader that the standard arguments have some form of `weak gravity' built in.  For instance, \cite{Yurtsever:2003ii} derives an area law for quantum fields in a box by putting a cutoff on the energy, but this automatically assumes that the binding energy is weak.  Likewise, \cite{Bousso:2014sda} provides a proof of the covariant entropy bound, again under the assumption of `weak gravity'.   Finally, the bound may be proven for classical systems under certain positive energy conditions \cite{Flanagan:1999jp}. However, in the quantum theory one must consider the binding energy of matter which is negative for mutually attractive particles.  In contrast, for super-extremal particles, electric repulsion naturally pushes particles to `weak gravity' regime.   In some sense, one may thus view the Weak Gravity Conjecture as a means of enforcing the positive energy / weak gravity conditions necessary for the proofs of \cite{Flanagan:1999jp}.

In Section 2 we will review the microscopic intuition to support the statement that a violation of WGC entails a remnant problem, and thus an infinite {\it{microcanonical}} entropy in violation of various entropy bounds.  This section does not contain original work, rather, we merely review some well-known facts in light of  modern entropy bounds.  The bulk of our work is in Section 3 where we compare this analysis with a `macroscopic' description of bound states in terms of extremal black holes.   We incorporate the effect of the charged particle spectrum by computing the one-loop correction to the black hole {\it{entanglement}} entropy.

We first describe corrections from several types of {\it neutral} particles that have previously been obtained in the literature. The standard lore is that massive particles contribute only subleading corrections (in a curvature expansion) to the classical entropy and geometry of black holes.  The intuition behind this is very simple; loop corrections of massive particles generate corrections that may be expanded as a series in $\text{Energy}/\text{Mass}$, which are negligible in the IR (see~\cite{Kaloper:2015jcz, Goon:2016mil} for recent discussions).  

However, to properly address issues related to the Weak Gravity Conjecture it is essential that one consider {\it charged} particles.  The story becomes somewhat more involved in this case. It is intuitively clear that charged particles can have a significant impact on the black hole since superextremal particles, independently of how massive, allow the black hole to decay by Schwinger pair production.  Thus, superextremal particles will necessarily generate a large correction to the geometry. When super-extremal particles are not present in the spectrum (the situation forbidden by the WGC) such a black hole decay is not possible.  It is interesting to consider more carefully the borderline case of extremal particles which are at the threshold of destabilizing the black hole.  In fact, we will show more generally that loops of charged fermions (whether extremal or not)  provide some evidence for the WGC cutoff, $\Lambda \lesssim q M_{P}$.  However, we do not in general find agreement between between the one-loop correction to the entropy and the microscopic intuition of Section 2.

In section 4, we briefly discuss how the corrections to the entropy are modified in theories with multiple (possibly dyonic) charges and supersymmetry.  Finally, in the conclusion, we discuss the discrepancy between the loop-corrected entanglement entropy and the microscopic expectations.  Given the semi-classical arguments implying unbounded microcanonical entropy corrections when WGC is violated, we interpret this mismatch as a failure to incorporate entropy stored on the timelike singularities in the Reissner-Nordstrom background.  This would then give a consistent, though speculative, picture in which the entropy of WGC violating theories violates the Beckenstein bound.\footnote{Of course, this entire argument is {\it{reductio ad absurdum}}, so we do not mean to imply that this scenario really occurs in a UV complete theory!}

\section{Microphysical Intuition}

We would first like to discuss the microcanonical entropy of bound states in a theory which violates the Weak Gravity Conjecture.  For the sake of simplicity we will assume that there is at least one charged particle since this assumption may already be motivated by other means \cite{Banks:2010zn, Harlow:2015lma}.  We will also only consider the case of a single $U(1)$ for simplicity.

At the classical level, the entropy of bound states in the microcanonical ensemble was determined in the seminal work \cite{antonov} (see also\cite{1989ApJS}).  The initially surprising conclusion is that the entropy is not bounded for a sufficiently large number of particles, rather, it is monotonically increasing with the binding energy of the system.   This is known as the `gravothermal catastrophe' and it guarantees that  a system of point particles never comes to a statistical equilibrium characterized by ergodic behavior.  

One might hope that the situation is ameliorated in quantum mechanics though the improvement is only marginal.  In fact, one can show \cite{Jetzer1992163} that for a bound state of $N$ non-relativistic particles interacting via a $1/r$ potential under the combined Coulombic + Newtonian forces the ground state has a binding energy of order $N^{3}  \Delta m^{5}/M_{P}^{4}$, where $\Delta m^{2} = m^{2} - 2 q^{2} M_{P}^{2}$ is the `effective mass squared'.  As $m$ or $N$ is increased the binding energy increases rapidly, and with it the entropy as a function of the energy, since more states are pushed below any given energy threshold.  The problem is even worse for subextremal {\textit{relativistic}} bosons - one can show that there is actually no ground state whenever the number of particles, $N$ exceeds a critical value, $N_{crit} \sim M_{P}^{2}/\Delta m^{2}$.  In other words, there are states of arbitrarily negative energy in quantum mechanics.  A corollary of this is that the entropy at fixed energy for $N> N_{crit}$ is infinite, since we may form infinitely many states by adding kinetic energy to states with arbitrarily negative energy \footnote{Another route by which one may seemingly deduce a large entropy is to count field configurations in a finite volume, in which case the entropy would seemingly go like the volume.  However, this is not counting the number of bound states and so it is not representing the entropy of some well-defined system that will be confined to a particular region for a long period of time.}.

For systems of sub-extremal fermions interacting via a Newtonian potential, it is also well known that beyond a critical mass there will be no minimal energy static solution; this is just the well-known Chandrasekhar limit.  One would again formally expect divergent entropy because the infinitely negative gravitational potential can be compensated in an infinite number of ways by kinetic energy to produce a finite energy. 

We emphasize that in order for the reasoning presented above to be valid in practice the particles should carry a conserved charge; otherwise they may annihilate or decay and the information could be carried away as photons.  Thus, these considerations do not imply that a bound state of neutrons could have infinite energy since we expect that neutrons should be able to decay in UV embeddings of the standard model.  

The standard response to the issues raised above is that the system will undergo `gravitational collapse' and that a horizon will form shielding the outside observer from the problems within.   While this is true, it is entirely beside the point.  Invoking a horizon to shield away the bound state physics is tantamount to declaring that `ignorance is bliss' and renouncing any connection between the Beckenstein-Hawking entropy and statistics.  In this case, it would seem to be impossible to reconcile black hole physics with quantum mechanics.  A more satisfying response to these issues would at the very least require an explanation as to how the information content of the microscopic description is so dramatically reduced.  The difficulty of this problem is the essence of the information paradox.    

Our view is that in order to have a UV completion in a theory of quantum gravity we should require that the microscopic description of the bound state reproduce the Beckenstein-Hawking  or Wald entropy.  This is a very stringent criterion on the theory and it should be no surprise that a simple model of subextremal particles should fail this test.  The point we are making is that, at least naively, the statistical entropy can easily be made to exceed the Beckenstein bound, making it unlikely that the theory can be fixed while only adding subextremal fields. 

This situation with subextremal particles may be contrasted with the extremal and superextremal cases.  In the superextremal case there are no bound states and so no bound state entropy to consider.\footnote{One may still put such particles in a box at finite temperature, which is an interesting test of entropy bounds \cite{Bousso:2003cm}.} More interestingly, there is good evidence that in certain models bound states of extremal particles actually reproduce the macroscopic dynamics of black holes \cite{Morita:2013wfa,Gaiotto:2004ij}.  In \cite{Morita:2013wfa} for instance, it was shown via standard thermodynamic arguments that the behavior of virialized extremal p-branes interacting via p-form fields and dilaton-gravity parametrically reproduces the black p-brane solution, including the Beckenstein-Hawking entropy.  This is known as the `p-soup' model.   We find the result to be encouraging since it shows that the naive dynamics of interacting extremal particles can be sufficient to provide a microscopic model of the black hole entropy without making additional unnecessary assumptions.  If we now perturb the p-soup model by adding a small attraction, we would naturally expect the energy of each state to decrease, leading to an increase of entropy at each energy level and thus a violation of the Beckenstein-Hawking bound.  

It is very difficult to formally carry the discussion on microscopic bound state entropy of subextremal particles much further.  In fact, if the Weak Gravity Conjecture is true then these systems really have no consistent UV embedding and so it may well be impossible to make sense of such bound states in the end.\footnote{At the very least, one would need a UV regulator for systems of sub-extremal particles.  However, a short distance repulsion seems unlikely to prevent a violation of the CEB since the entropy may be understood as the result of the long distance attraction which reduces the ground state energy. For a large system of particles, the repulsion affects the total energy according to the number of nearest neighbors (i.e., $\sim N$) while the negative attraction energy goes like $\sim N^{3}$ \cite{Jetzer1992163}.  Thus, the long range force wins in the end.}  However, our strategy is to just take these bound states at face value and draw what conclusions we may.   We find the simple arguments made here to be very suggestive and see no easy way to evade the `catastrophe' pointed out long ago \cite{antonov}.  In other words, we are claiming that a theory of mutually attractive point particles carrying a conserved charge will not satisfy the usual holographic entropy bounds.  

As an indirect test of the entropic contribution one gets from different kinds of particles, in the next section we compute the change in entanglement entropy of an extremal Reissner-Nordstrom black hole.  It is not at all a-priori clear how this entanglement entropy should relate to the microcanonical entropy discussed above and the discrepancy we find will be discussed in the following section.

\section{Macroscopic Description}

The microscopic arguments described above hint that {\it subextremal} particles should give rise to bound states with large, possibly divergent entropy while {\it superextremal} particles exhibit no such problems. Furthermore, in the `p-soup' model of {\it extremal} particles, the Bekenstein-Hawking bound seems to be parametrically satisfied, although it is not a priori clear that this will be the case for other models of extremal particles.  We would like to ask whether there are any hints of this kind of behavior in the gravitational description of such bound states as an extremal black hole.  In particular, do extremal black holes in theories with only subextremal particles exhibit divergent entropy?  What about when there are only extremal particles?

In order to incorporate the effect of the charged particle spectrum we will compute the one loop corrected black hole geometry and entropy using the quantum entropy function formalism of \cite{Sen:2007qy, Banerjee:2010qc}. The goal is to obtain the leading corrections to the Beckenstein-Hawking entropy.

Integrating out charged particles induces an infinite series of curvature and field strength corrections. We may then calculate the entropy by applying the Beckenstein-Hawking-Wald formula \cite{Wald:1993nt} to the effective action for the massless fields. This entropy may also be formally related to the entanglement entropy of the horizon \cite{Solodukhin:2011gn}.  

We emphasize that this computation does not directly give us the microcanonical entropy that we are ultimately after; nevertheless, we may still hope to uncover some pathologies via this exercise. Independently of WGC, this calculation may have some interest in its own right for the purpose of precision matching between macroscopic and microscopic entropy calculations, though we will not pursue this here.

Finally, a word of caution before we begin.  When we integrate out a subextremal particle we are implicitly assuming that the EFT cutoff is well above the particle mass.  On the other hand, the electromagnetic dual of the WGC condition $m < \sqrt{2} q M_{P}$ implies that the cutoff obeys $\Lambda \lesssim q M_{P}$, which is the so called `magnetic' version of the WGC \cite{ArkaniHamed:2006dz}.  Such a cutoff would push subextremal particles outside the realm of effective field theory and so they should really be thought of as solitons from the low energy perspective.  Thus, our calculation is simultaneously assuming a violation of both the electric and magnetic forms of the conjecture.  The issues that we raise could therefore be interpreted as tracing back to either, or both assumptions.

\subsection{Classical entropy function}
We will focus for the most part of this section on electrically charged extremal black holes of charge $Q$, in the simplest setup of Einstein-Maxwell theory in 3+1 dimensions. We will study the effects of matter fields (either scalars, fermions, or an ${\mathcal N}=1$ chiral multiplet) of charge $q$ and mass $m$.  We take the action to be:  
\beq
\label{sgrem}
S &=& \int d^{4}x\sqrt{-g}  \left(\mathcal{L}^{(0)}_{GR+EM} + \mathcal{L}_{matter}\right) \\ \nonumber
\\  \nonumber
{\mathcal L}^{(0)}_{GR+EM} &=&\frac{M_{P}^{2}}{2} \mathcal{R} -\frac{1}{4} F_{\mu\nu}F^{\mu\nu} \\ \nonumber
\\ \nonumber
{\mathcal L}_{matter} &=&
\begin{cases}
      {\mathcal L}_{scalar}=-(D_{\mu}\phi)^{\star}D^{\mu}\phi-m^{2}|\phi|^{2} ,\\
      {\mathcal L}_{fermion}=\overline{\psi} \left(\slashed{D}-m\right) \psi,\\
      {\mathcal L}_{{\cal N}=1}={\mathcal L}_{scalar}+{\mathcal L}_{fermion}\,.
    \end{cases} \\ \nonumber
\eeq

Integrating out the matter fields will induce a 1-loop correction so that the total effective action for the massless fields will be of the form $\mathcal{L}_{GR+EM} = \mathcal{L}_{GR+EM}^{(0)}+\mathcal{L}_{GR+EM}^{(1)}$.  The correction term may be computed using the heat-kernel formalism \cite{Vassilevich:2003xt}.  We are interested in how black holes in this 1-loop corrected theory depend upon the parameters $q$ and $m$.  In particular, we are interested in computing the entropy as a function of the charge measured at infinity.

It is a useful observation that regardless of the detailed form of the corrections, the near-horizon geometry is described by $AdS_{2}\times S^{2}$, which we write as:
\beq
\label{bg}
ds^{2}&\equiv& g_{\mu\nu}dx^{\mu}dx^{\nu} = a^2\left(-r^{2} dt^{2}+\frac{dr^{2}}{r^{2}}\right) + b^2\left(d\theta^{2} +\sin^{2}\theta d\phi^{2}\right) \\ \nonumber
F &=& E dt\wedge dr
\eeq
Here $a$ and $b$ parameterize the radii of $AdS_2$ and $S^2$, respectively, and $E$ represents the electric field sourced by the black hole. In terms of this parameterization, the Wald entropy is now given by \cite{Sen:2007qy} minimizing the entropy functional, $\mathcal{E}$,  defined by:
\be
\label{ES}
\mathcal{E}(Q;E,a,b) = 2 \pi\left[Q E - 4\pi a^2\, b^2 \, \mathcal{L}_{GR+EM}(E,a,b)\right]\,,
\ee
with $\mathcal{L}_{GR+EM}$ evaluated on the near-horizon geometry~\eqref{bg}. More precisely, the entropy of a black hole of electric charge $Q$ is given by $\mathcal{S}(Q) =\underset{a,b,E}{\min}\, \mathcal{E}$.  

At tree level the entropy is straightforward to compute.  First, the lagrangian density is computed on the background (\ref{bg}).  One finds:
\be
\label{lag}
\mathcal{L}_{GR+EM}^{(0)} = M_{P}^{2}\left(\frac{1}{b^2}- \frac{1}{a^2}\right)+\frac{E^{2}}{2 a^{4}}
\ee
Plugging this into the expression for $\mathcal{E}$ and minimizing we immediately get the equations:
\beq
\label{tree}
E_0 &=& \frac{Q}{4\pi} \\ \nonumber
a_0^2 &=& b_0^2 = \frac{Q^{2}}{32 \pi^{2} M_{P}^{2}} \\ \nonumber
\eeq
For the sake of reference, the ADM mass of the full solution is also known to be $M = \sqrt{2} Q M_{P}$.  Finally, plugging this solution back in to (\ref{ES}) we find the expected Beckenstein-Hawking formula:
\be
\mathcal{S}^{(0)} = \frac{Q^{2}}{4} = \frac{A}{4 G_{N}}
\ee

\subsection{One-Loop Correction}

We now wish to repeat this procedure with the 1-loop corrections induced by the matter fields included.  In general, these corrections may be calculated using the heat kernel formalism, as explained in \cite{Banerjee:2010qc}.    To summarize, one first computes the heat kernel, $K(x,y;s)$, defined by:
\be 
(\partial_{s}-D)K(x,y;s) = 0 \,\,\,\,\,\,\,\,\,\,\,\,\,  K(x,y;0) = \delta^{4}(x-y)
\ee
where $D$ is a generalized laplacian containing the kinetic and mass terms of the field being integrated out.  Once this is known, the one loop correction is:
\be
\label{1loop}
\mathcal{L}^{(1)} =\frac{1}{2} \int^{\infty}_{\epsilon^{2}} \frac{ds}{s} K(s)
\ee
where $\epsilon$ is a UV cutoff.  The notation $K(s)$ is shorthand for $K(x,x;s)$ which is independent of $x$ in the near-horizon geometry. In the present case, the geometry is the product space $AdS_{2}\times S^{2}$ and so the heat kernel factorizes into $K = K_{AdS_{2}}\times K_{S^{2}}$.  This fact allows one to compute the heat kernel using the techniques of \cite{Comtet:1984mm,Pioline:2005pf, Banerjee:2010qc}.  The total $K$ is just the sum of the result for the different matter fields considered. The results are as follows:
\subsubsection*{Charged Scalar}
\beq
\label{Kb}
K_{s}(s)&=& \frac{e^{-s \Delta m^{2}}}{4\pi^{2}a^2 b^2 }\sum_{l=0}^{\infty}(2l +1)\int_{0}^{\infty} d\lambda\, \lambda\,\rho_s(\lambda)e^{-s\left[\left(\lambda^{2}+\frac{1}{4}\right)/a^2+l(l+1)/b^2\right]} \\  \nonumber
\rho_s(\lambda) &=& \frac{\sinh(2\pi \lambda)}{\cosh (2\pi \lambda)+\cosh(2\pi q E)} \\ \nonumber
\eeq
\subsubsection*{Chiral Fermion}
\beq
\label{Kf}
K_{f}(s)&=& \frac{e^{-s \Delta m^{2}}}{4\pi^{2} a^2 b^2} \sum_{l=0}^{\infty} (2 l +2) \int_{0}^{\infty} d\lambda \,\lambda\, \rho_{f}(\lambda) \,e^{-s\left[\lambda^{2}/a^2 + (l+1)^{2}/b^2\right]} \\ \nonumber
\rho_{f}(\lambda)&=& \frac{\sinh(2\pi \lambda)}{\cosh(2\pi q E)-\cosh(2\pi \lambda)}
\eeq

Here, $\lambda$ labels a momentum mode in $AdS_{2}$ with physical momentum $\lambda/a$ and $l$ labels an angular momentum mode in $S^{2}$.  The functions $\rho_{s}(\lambda)$ and $\rho_{f}(\lambda)$ are the spectral densities for bosons and fermions respectively  (for more details on the nature of the solution, see \cite{Comtet:1984mm,Pioline:2005pf}).  Finally,  the heat kernel for an ${\mathcal N}=1$ chiral multiplet is simply obtained by adding these two expressions: ${K}_{{\cal N}=1}=K_{s}+ K_{f}$.

We have introduced the notation $\Delta m^{2}$ which will be critical for the rest of our discussion.  In general, this takes the form:
\be
\Delta m^{2} = m^{2} - \frac{q^{2} E^{2}}{a^2}
\ee
If the classical description is valid then by (\ref{tree}) this becomes:
\be
\Delta m^2 = m^{2} - 2 q^{2} M_{P}^{2}
\ee
Therefore, we see that to leading order $\Delta m^{2}$ precisely delineates the extremality bound that we would infer from large black holes.  We will henceforth refer to $m^{2} > 2 q^{2} M_{P}^{2}$, $m^{2} < 2 q^{2} M_{P}^{2}$ and  $m^{2}=2 q^{2} M_{P}^{2}$ as sub-, super-, and extremal particles, respectively.  

It is obvious from equations  (\ref{Kb}) and (\ref{Kf}) that the convergence of the $s$ integral in (\ref{1loop}) will depend on the sign of $\Delta m^{2}$.  In particular, for subextremal particles (such that $\Delta m^2\geq 0$), the $s$ integral will always be suppressed at large $s$.  On the other hand, superextremal particles (such that $\Delta m^2< 0$) give rise to IR divergences for sufficiently large black holes.  The extremal case ($\Delta m^{2}=0$) requires special care since the suppression of the $s$ integral comes from the factor $e^{-s/(4a^{2})}$ (bosons) or $e^{-s/b^{2}}$ (fermions).   Given the importance of the $e^{-s\Delta m^2}$ exponential suppression, we find it convenient to separate this out explicitly.  We will thus introduce the following notation to facilitate discussion:
\be
\label{KKtilde}
K(s) \equiv e^{-s \Delta m^{2}} \tilde{K}(s)
\ee
We will study the heat kernels in more detail in Sections~\ref{sec:neutral} and~\ref{sec:kernels}.

The purpose of the remainder of this section is to use the formulae presented above to compute the corrections induced by super-, sub-, and extremal particles. To illustrate the methods and the type of results we would a priori expect,  we first reproduce the computation of corrections induced by neutral particles~\cite{Banerjee:2010qc}. 

\subsection{Neutral particles: $\Delta m^2=m^2$}\label{sec:neutral}
\subsection*{The neutral heat kernels}
The heat kernels for neutral scalars and fermions can be obtained by setting $q=0$ in eqs.~\eqref{Kb} and~\eqref{Kf}, which amounts to replacing $\Delta m^2 \to m^2$, and 
\be
\rho_{s}(\lambda)\to \tanh(\pi \lambda)\,,\qquad\qquad \rho_{f}(\lambda)\to -\coth(\pi \lambda)
\ee

We are interested in the leading contributions to the entropy in a large $Q$ (equivalently large $a$) expansion. These arise from the region $s\ll a^{2}\sim b^{2}$ in the integral~\eqref{1loop}, and so we need  the small-$s$ expansion of $\tilde{K}(s)$. Care is needed in obtaining this series so that the divergent terms are properly isolated (see \cite{Banerjee:2010qc}). The $\lambda$-dependent contributions from the $AdS_2$ factor, and the $l$-dependent ones from the $S^2$ can be treated separatedly.
One can expand $\rho_s(\lambda)$, the density of states in $AdS_2$ as:
\be
\rho_s(\lambda)=\tanh(\pi\lambda)=\frac{1-e^{-2\pi\lambda}}{1+e^{-2\pi\lambda}}=1+2\sum_{k=1}^{\infty}(-1)^ke^{-2\pi k\lambda}
\ee
We may use this expansion to analytically perform the $\lambda$ integral in the heat kernel~\eqref{Kb} (for convenience, we introduce the notation $\bar{s}\equiv s/a^2$, we look hence for an expansion in $\bar{s}\ll 1$):
\begin{eqnarray}\label{neutralKAdS}
\int_0^\infty d\lambda\,\lambda \,\rho_s(\lambda)e^{-\bar{s}\lambda^2}&=&\int_0^\infty d\lambda\, \lambda\, e^{-\bar{s}\lambda^2}\left(1+2\sum_{k=1}^\infty (-1)^k e^{-2\pi k\lambda}\right)\nonumber\\
&=&\frac{1}{2\bar{s}}+2\sum_{n=0}^\infty\frac{(-\bar{s})^n}{n!}\sum_{k=1}^{\infty}(-1)^k\int_0^\infty d\lambda\, \lambda^{2n+1}e^{-2\pi k \lambda}\nonumber\\
&=& \frac{1}{2\bar{s}}+2\sum_{n=0}^\infty\frac{(-\bar{s})^n}{n!}\frac{(2n+1)!}{(2\pi)^{2n+2}}\zeta(2n+2)\left(2^{-2n-1}-1\right)\nonumber\\
&=&\frac{1}{2\bar{s}}-\frac{1}{24}+\frac{7}{960}\bar{s}+{\cal{O}}(\bar{s}^2)
\end{eqnarray}

A similar method may be used to obtain the contribution from the $S^2$ factor, i.e. to perform the summation over $l$ in~\eqref{Kb}.  The first trick is to rewrite the sum as an integral analogous to the one analyzed above, using the contour integration identity (now defining $\tilde{s}\equiv s/b^2$):
\be
\label{suml}
\sum_{l=0}^{\infty} (2l + 1)e^{-\tilde{s} (l+1/2)^{2}} = - i \oint d\lambda \lambda \tan(\pi \lambda) e^{- \tilde{s}\lambda^{2}}
\ee
where the contour is chosen to encircle the positive real axis.  Now we have to deal with the integral of $\tan$.  Note, however, that upon deforming the contour to the imaginary axis this becomes $\tanh$, which is precisely the same oject we evaluated above.  Therefore, (\ref{suml}) may be evaluated order by order in $\tilde{s}$, just as we did above in~\eqref{neutralKAdS}:
\be\label{neutralKS}
\sum_{l=0}^{\infty} (2l + 1)e^{-\tilde{s} (l+1/2)^{2}} =\frac{1}{\tilde{s}}+\frac{1}{12}+\frac{7}{480}\tilde{s}+{\cal O}(\tilde{s}^2)
\ee

The total heat kernel for scalars, in a convenient large radius expansion is obtained by plugging~\eqref{neutralKAdS} and~\eqref{neutralKS} into~\eqref{Kb}:
\be\label{neutralexpansions}
{K}_s(s)=\frac{e^{-sm^2}}{8\pi^2 s^2}\left[1-s\left(\frac{1}{3a^2}-\frac{1}{3b^2}\right)+s^2\left(\frac{1}{15a^4}+\frac{1}{15b^4}-\frac{1}{9a^2b^2}\right)+{\mathcal O}(s^3)\right]
\ee

A similar calculation can be performed for the fermionic case, with the result
\be\label{neutralexpansionf}
K_f(s)=\frac{-e^{-sm^2}}{8\pi^2 s^2}\left[1+s\left(\frac{1}{6a^2}-\frac{1}{6b^2}\right)-s^2\left(\frac{1}{60a^4}+\frac{1}{60b^4}+\frac{1}{36a^2b^2}\right)+{\mathcal O}(s^3)\right]
\ee
These expressions are valid in the region $\epsilon^2<s\ll a^{2}\sim b^{2}$. It is straightforward to integrate them to obtain the leading contributions to ${\mathcal L}^{(1)}$, which can subsequently be used to compute the quantum corrected entropy functional ${\mathcal E}$ and the black hole entropy ${\mathcal S}(Q)$. The cases of massless and massive particles must be handled separately:
\subsubsection*{Massive neutral particles}
When $m^2 \gg a^{-2}$, the exponential damping induced by the $e^{-sm^2}$ term in $K(s)$ suppresses the contributions from the region $m^2\gg s^{-1}$,  both in the near horizon geometry and in flat space. The (not unexpected) conclusion is that massive particles do not induce corrections that scale logarithmically with the black hole charge~\cite{Banerjee:2010qc}. This can be explicitely checked from the one-loop effective Lagrangian:
\begin{eqnarray}\label{deltaL}
{\cal L}^{(1)}_s &=&  \frac{1}{32\pi^2 \epsilon^4} -\frac{m^2}{16\pi^2\epsilon^2}+\frac{1}{48\pi^2 \epsilon^2}\left(\frac{1}{b^2}-\frac{1}{a^2}\right) \nonumber\\
& -&\left[\frac{m^4}{32\pi^2}  -\frac{m^2}{48 \pi^2}\left(\frac{1}{b^2}-\frac{1}{a^2}\right) +\frac{1}{240\pi^2}\left(\frac{1}{a^4}+\frac{1}{b^4}\right)-\frac{1}{144\pi^2 a^2 b^2} \right]\ln{(\epsilon^2 \, m^2})\nonumber
\end{eqnarray}
The UV divergences parameterized by $\epsilon$ are reabsorbed in renormalized coupling constants, exactly canceling the logarithmic contribution. The quantum corrected entropy function takes exactly the same form as the classical one, although expressed in terms of one-loop renormalized coupling constants:
\begin{equation}
{\cal L}_s^{(0)}+{\cal L}_s^{(1)}=M_P^2\left(\frac{1}{b^2}-\frac{1}{a^2}\right)+\frac{E^2}{2a^4}
\end{equation}
Hence, massive neutral particles do not induce logarithmic corrections to the black hole entropy. A similar conclusion follows straightforwardly for fermions.
\subsubsection*{Massless neutral particles}
This case is more interesting, since it leads to entropy corrections that scale logarithmically with the black hole charge. The crucial point is to notice that, even if the field is massless, there is a mass gap in the spectrum in the near horizon geometry, induced by the curvatures of $AdS_2$ and $S^2$. This leads to an exponential IR suppression in the heat kernel for large $s$ (c.f.~\eqref{Kb} and~\eqref{Kf}), that yields terms in the effective Lagrangian proportional to $\log(a/\epsilon)$. This is unlike in flat space, where the spectrum is gapless, and there is no such exponential suppression of the heat kernel.

We can see how this works explicitly by using~\eqref{1loop} and the heat kernel expansion~\eqref{neutralexpansions} and~\eqref{neutralexpansionf}:
\be\label{L1neutrals}
{\mathcal L}^{(1)}_s=\frac{1}{32\pi^2 \epsilon^4}-\frac{1}{48\pi^2 \epsilon^2}\left(\frac{1}{a^2}-\frac{1}{b^2}\right)+\frac{1}{48\pi^2}\left(\frac{1}{5a^4}+\frac{1}{5b^4}-\frac{1}{3a^2b^2}\right)\log\left(\frac{a^2}{\epsilon^2}\right)
\ee
\be\label{L1neutralf}
{\mathcal L}^{(1)}_f=-\frac{1}{32	\pi^2 \epsilon^4}-\frac{1}{96\pi^2 \epsilon^2}\left(\frac{1}{a^2}-\frac{1}{b^2}\right)+\frac{1}{192\pi^2}\left(\frac{1}{5a^4}+\frac{1}{5b^4}+\frac{1}{3a^2b^2}\right)\log\left(\frac{b^2}{\epsilon^2}\right)
\ee
Again, UV divergences appear and can be absorbed into the renormalized coupling constants (see Appendix). In particular, the quartic divergences arise from the renormalization of the cosmological constant, quadratic ones from Newton's constant, and logarithmic divergences from higher curvature terms. By combining~\eqref{L1neutrals} and~\eqref{L1neutralf} with the tree level lagrangian~\eqref{tree}, and comparing with the renormalized quantities in~\eqref{renormalized}, we can express the total Lagrangian, and hence the entropy function as:\footnote{Here, and in the following, we are setting the renormalized cosmological constant to zero $\Lambda^{(r)}=0$, since we are interested in asymptotically flat spacetimes.}
\begin{eqnarray}
&&{\mathcal E}_s(Q;E,a,b) = 2 \pi\left[Q E - 4\pi a^2\, b^2 \,\left( \mathcal{L}^{(0)}+\mathcal{L}^{(1)}\right)\right]\\
&&= 2 \pi\left\{Q E - 4\pi a^2\, b^2\,\left[M_P^2\left(\frac{1}{b^2}-\frac{1}{a^2}\right)+\frac{E^2}{2a^4}+\frac{1}{48\pi^2}\left(\frac{1}{5a^4}+\frac{1}{5b^4}-\frac{1}{3a^2b^2}\right)\log\left(a^2 \mu^2\right)\right]\right\}\nonumber
\end{eqnarray}
\begin{eqnarray}
&&{\mathcal E}_f(Q;E,a,b) = 2 \pi\left[Q E - 4\pi a^2\, b^2 \,\left( \mathcal{L}^{(0)}+\mathcal{L}^{(1)}\right)\right]\\
&&= 2 \pi\left\{Q E - 4\pi a^2\, b^2\,\left[M_P^2\left(\frac{1}{b^2}-\frac{1}{a^2}\right)+\frac{E^2}{2a^4}+\frac{1}{192\pi^2}\left(\frac{1}{5a^4}+\frac{1}{5b^4}+\frac{1}{3a^2b^2}\right)\log\left(b^2\mu^2\right)\right]\right\}\nonumber
\end{eqnarray}
where $\mu$ is a scale that must be introduced in the renormalization procedure in flat space when integrating out massless particles (see Appendix). We are interested in the scaling of the entropy with $a$ (or equivalently $Q$), so the scale $\mu$ plays no role in our discussion. 

One may now extremize ${\cal E}_s$ and ${\cal E}_f$ to obtain the corresponding quantum corrected black hole entropies. It is simpler, however, to substitute the classical parameters $E_0$, $a_0$, and $b_0$ of~\eqref{tree} to obtain a quick estimate of the correction:\footnote{Note that the correctons below are negative, which is actually a fairly general phenomenon.  Negative entropy has primarily been discussed in the context of gauge fields \cite{Kabat:1995eq, Donnelly:2012st} and have also been found in many stringy black hole backgrounds as reviewed by \cite{Sen:2012dw}.  In fact, it has been conjectured by Giddings \cite{Giddings:2013vda} that negative corrections to the black hole entropy are actually necessary in order to solve the information paradox.}
\be\label{neutralSs}
{\mathcal S}_s(Q)\approx {\mathcal E}_s(Q;E_0,a_0,b_0)=\frac{Q^2}{4}-\frac{1}{90}\log Q^2
\ee
\be\label{neutralSf}
{\mathcal S}_f(Q)\approx {\mathcal E}_f(Q;E_0,a_0,b_0)=\frac{Q^2}{4}-\frac{11}{360}\log Q^2
\ee
To this order, the classical values $E_0$, $a_0$ and $b_0$, represent the extremum of the quantum corrected entropy function and so~\eqref{neutralSs} and~~\eqref{neutralSf} give the exact logarithmic corrections to the black hole entropy induced by massless neutral particles~\cite{Banerjee:2010qc}. 

One can also compute the entropy corrections from a massless ${\cal N}=1$ chiral multiplet, by just adding the contributions to the lagrangian of a scalar and a fermion, with the result $\Delta {\cal S}_{{\cal N}=1}(Q)=-\frac{1}{24}\log(Q^2)$.  These results have been generalized to stringy setups and matched against microscopic entropy computations, finding exact agreement~\cite{Banerjee:2010qc}.  

We see from these examples that massless and massive neutral particles have qualitatively different effects on the entropy. However, in both cases, the classical solution is a good approximation in a large charge expansion, and corrections are subleading with respect to the classical entropy, starting at log(Q).  We now want to generalize these computations to the case of charged particles. 

\subsection{Superextremal Particles $(m^2<2q^2 M_P^2)$: Unstable black holes}

%
As mentioned previously, when $\Delta m^2<0$, the spectrum contains tachyonic modes and the $s$ integral in~\eqref{1loop} is IR divergent, signaling the instability of the near-horizon geometry.\footnote{Strictly speaking, the IR divergence appears when $\Delta m^2<\frac{-1}{4a^2}$ for bosons and when $\Delta m^2<\frac{-1}{b^2}$ for fermions. For sufficiently large black holes, the condition reduces to $\Delta m^2 <0$ in both cases. This hints that the WGC must be phrased in terms of a strict inequality in the Einstein-Maxwell theory as also suggested in \cite{Ooguri:2016pdq}. This is also related to the Breitenlohner-Freedman bound of the $AdS_2$ with radius $a$ for the bosons, while for the fermions it seems to be related to the $S^2$ of radius $b$.} This is not unexpected since, in the presence of super-extremal particles, Schwinger pair production occurs near the horizon and leads to the emission of charged particles and the discharge of the black hole. In fact, by careful analytic continuation of the above equations, one can estimate the rate of emission of super-extremal particles from an extremal black hole~\cite{Pioline:2005pf}.

\subsection{The heat kernel for charged particles}\label{sec:kernels}
Once the particles mass rises to the extremality bound, the Schwinger pair production decay channel shuts off and the black hole becomes exactly stable. We would like to perform calculations analogous to those of Section~\ref{sec:neutral} in this case. The first step requires us to obtain the large radius expansions (more precisely, the $s/a^2\ll1$ expansions) of the charged heat kernels~\eqref{Kb} and~\eqref{Kf} analogous to~\eqref{neutralexpansions} and~\eqref{neutralexpansionf}. This will prove to be a much more subtle task than in the neutral case.

\begin{figure}[htb]
\begin{subfigure}{.5\textwidth}
  \centering
  \includegraphics[width=.8\linewidth]{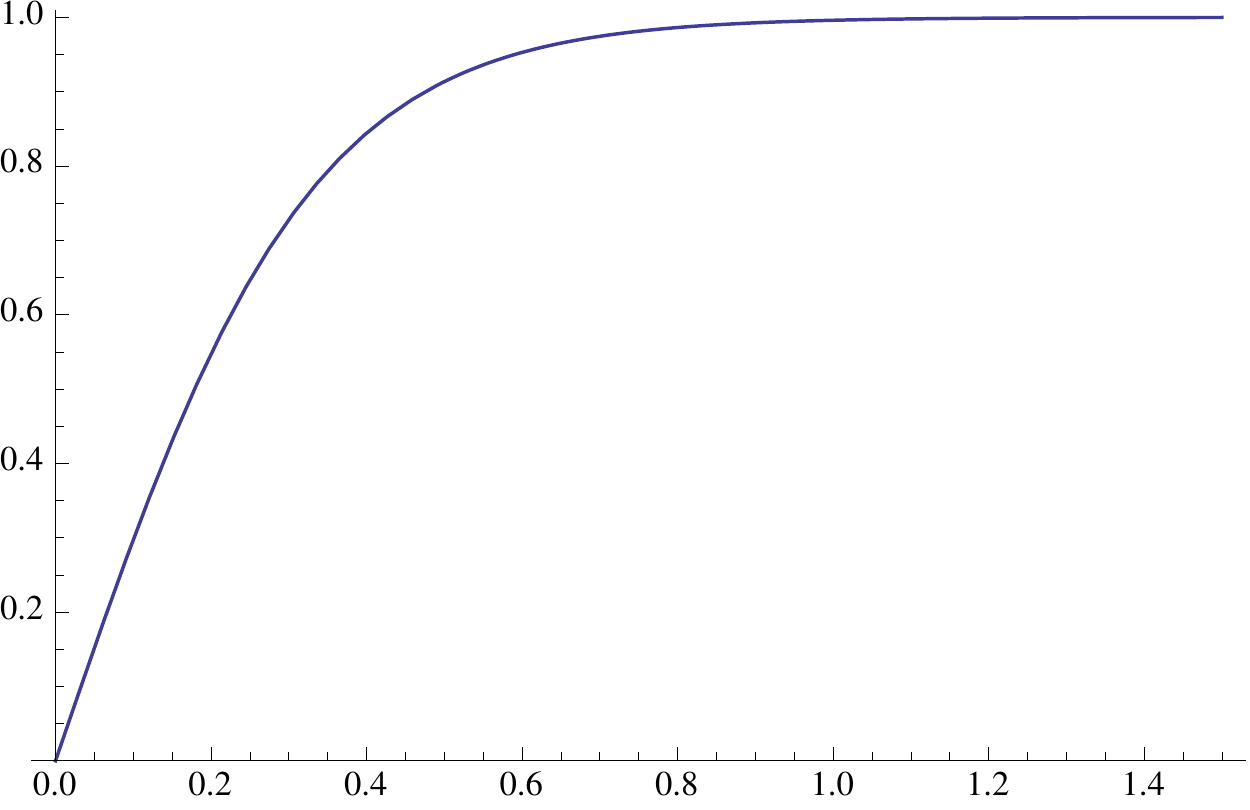}
  \caption{Intermediate Black Hole ($qE=0.1$)}
  \label{Intermediate}
\end{subfigure}%
\begin{subfigure}{.5\textwidth}
  \centering
  \includegraphics[width=.8\linewidth]{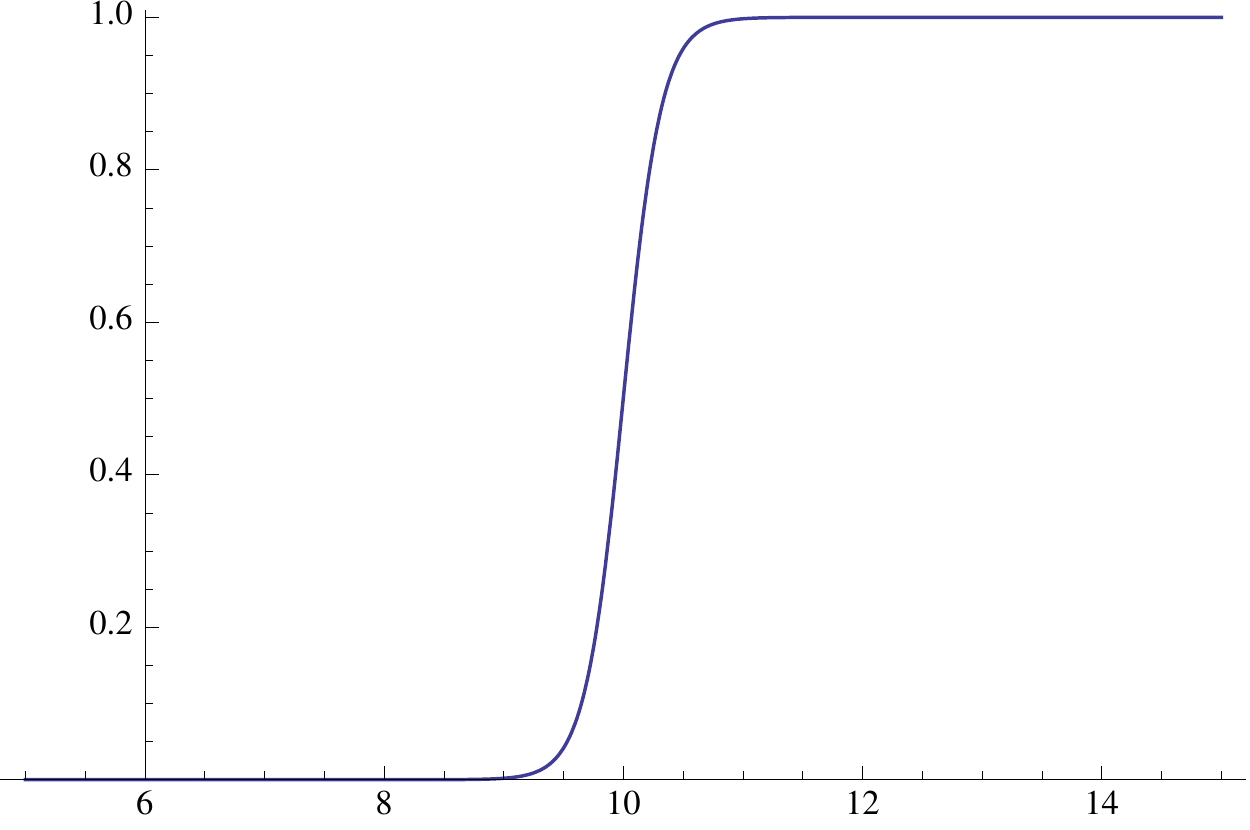}
  \caption{Large Black Hole ($qE=10$)}
  \label{Large}
\end{subfigure}
\caption{Spectral densities for large and small $q E$.  One can see that the spectral density for $q E \gg 1$ resembles a step function.}
\label{fig:densities}
\end{figure}

Let us focus first on the scalar heat kernel. The $l$-dependent contribution from the $S^2$ factor is unchanged with respect to the neutral case~\eqref{neutralKS}. The $\lambda$-dependent contribution from the $AdS_2$ sector changes significantly. Given its importance, we reproduce expression~\eqref{Kb} once again here:
\beq
\label{Kernelb}
K_{s}(s)&=& \frac{e^{-s \Delta m^{2}}}{4\pi^{2}a^2 b^2 }\sum_{l=0}^{\infty}(2l +1)\int_{0}^{\infty} d\lambda\, \lambda\,\rho_s(\lambda)e^{-s\left[\left(\lambda^{2}+\frac{1}{4}\right)/a^2+l(l+1)/b^2\right]} \\  \nonumber
\rho_s(\lambda) &=& \frac{\sinh(2\pi \lambda)}{\cosh (2\pi \lambda)+\cosh(2\pi q E)} \\ \nonumber
\eeq
Notice that, through the electric field $E$, the black hole radius now enters the spectral density $\rho_s(\lambda)$, since classically $E\sim a M_p$. Much care is needed in performing the appropriate expansion of this function. Two very different regimes can be distinguished: {\it large} black holes $qE\gg 1$ and {\it intermediate} black holes $qE\ll 1$. Figure~\ref{fig:densities} shows the different forms of the spectral densities in the two regimes. 

The physical interpretation of the two regimes is as follows. While in both of them the black hole radii remain above the Planck scale, and hence remain in a regime where effects of quantum gravity should be negligible,  {\it intermediate} black holes are those whose curvature (the inverse radius) lies above the WGC cutoff $\Lambda_{WGC}=qM_p$, that is: $qM_p\ll 1/a \ll M_p$. {\it Large} black holes are those whose curvature is smaller even than the WGC scale.  Another way of saying this is that intermediate black holes are also those for which probe extremal particles are effectively massless since the curvature coupling provides the more stringent IR cutoff, just as in the massless neutral case.   Finally, note that generically we can maintain a hierarchy between the black hole and particle mass since $Q M_{P} \gg q M_{P}$.

Given the qualitative different behavior of the spectral densities shown in Figure~\ref{fig:densities}, the corresponding expansions of the heat kernels will be necessarily different. In particular, the spectral density for intermediate black holes is very similar to that of neutral particles, and an expansion analogous to that used in section~\ref{sec:neutral} will be valid. As we shall see momentarily, this expansion will not be sufficient for large black holes. 

\subsection*{Scalar heat kernel for intermediate black holes $qE\ll 1$}
The first step is to develop an expansion for $\rho_{s}$ as a series in $e^{-2\pi \lambda}$.  One can check that formally:
\beq
\label{rhos}
\rho_{s}(\lambda)&=& \frac{1- e^{-4\pi\lambda}}{1+e^{-4\pi \lambda}+2 e^{-2\pi \lambda} \cosh(2\pi q E)} \\ \nonumber
&=&1+ 2\sum_{k=1}^{\infty} (-1)^{k} \cosh(2\pi k q E) e^{-2\pi k \lambda}
\eeq
While this expansion naively fails for $\lambda < q E$,  we will be able to exactly resum the $k$ expansion after performing the $\lambda$ integral at fixed $s$, thus circumventing this issue. To see how this is done, we now plug~\eqref{rhos} into the heat kernel expression~\eqref{Kernelb} and do the $\lambda$ integral analytically.  For convenience, define $\os= s/a^{2}$. One finds:
\beq\label{intermediateKAdS}
&&\int_{0}^{\infty} d\lambda\, \lambda \rho_{s}(\lambda) e^{-\os \lambda^{2}} = \int_{0}^{\infty} d\lambda \lambda\left(1+ 2\sum_{k=1}^{\infty} (-1)^{k} \cosh(2\pi k q E) e^{-2\pi k \lambda}\right)e^{-\os \lambda^{2}} \nonumber \\ \nonumber 
&=&\frac{1}{2\os} + 2\sum_{k=1}^{\infty}(-1)^{k} \cosh(2\pi k q E)\sum_{n=0}^{\infty}\frac{(-\os)^{n}}{n!} \int_{0}^{\infty} d\lambda \,\lambda^{2n+1} e^{-2\pi k \lambda} \\ \nonumber
&=&\frac{1}{2\bar{s}}+2\sum_{n=0}^\infty (-1)^n\frac{\bar{s}^n}{n!}\frac{(2n+1)!}{(2\pi)^{2n+2}} \sum_{k>0}\frac{(-1)^k}{k^{2n+2}}\cosh{(2\pi k q E)} \nonumber\\
&=&\frac{1}{2\bar{s}}+\frac{1}{2\pi^{2}} \sum_{k>0}\sum_{m\geq 0}\frac{(-1)^k}{k^{2}}\frac{(2\pi k q E)^{2m}}{(2m)!}-\frac{3\bar{s}}{4\pi^{4}} \sum_{k>0}\sum_{m\geq 0}\frac{(-1)^k}{k^{4}}\frac{(2\pi k q E)^{2m}}{(2m)!}
+{\cal O}(\bar{s}^2) \nonumber\\
&=&\frac{1}{2\bar{s}}+\frac{1}{4\pi^{2}} \sum_{m\geq 0}\frac{(2\pi q E )^{2m}}{(2m)!}(-2+2^{2m})\zeta(2-2m)  -
\frac{3\bar{s}}{32\pi^{4}} \sum_{m\geq 0}\frac{(2\pi q E)^{2m}}{(2m)!}(-8+2^{2m})\zeta(4-2m) +
{\cal O}(\bar{s}^2) \nonumber\\
&=&\frac{1}{\bar{s}}\left(\frac{1}{2}\right)-\left(\frac{1}{24}+\frac{(q E)^2}{2}\right) + \bar{s} \left( \frac{7}{960}+\frac{1}{8} q^{2} E^2 +\frac{1}{4}q^{4} E^4\right) + {\cal O}(\bar{s}^2) 
\eeq
In the second to last step we have recognized the $k$ summation as giving the zeta function and in the last step, we have used that $\zeta(-2k)=0$ for $k\in \mathbb{N}$, which implies that for each order in $s$ the $q$ expansion truncates at finite order.

The heat kernel~\eqref{Kernelb} for charged scalars in the background of an extremal intermediate black hole can be conveniently expressed in a large radius ($s\ll a^2\sim b^2$) and small $qE\ll 1$ expansion by combining~\eqref{intermediateKAdS} and~\eqref{neutralKS} :
\begin{eqnarray}\label{totalkernels}
K_{\text{medium},s}(s) &=&e^{-s\Delta m^2} \left\{\frac{1}{8\pi^2 s^2}+\frac{1}{24\pi^2 s}\left(\frac{1}{b^2}-\frac{1}{a^2}-\frac{3 q^2 E^2}{a^2}\right) \right. \\
&+&\left. \left[\frac{1}{120\pi^2}\left(\frac{1}{a^4}+\frac{1}{b^4}\right)-\frac{1}{72\pi^2 a^2 b^2} -\frac{q^2 E^2}{24\pi^2 a^2 b^2} +\frac{q^2 E^2}{16\pi^2 a^4} + \frac{q^4 E^4}{16\pi^2 a^4}\right] +{\cal O}(s)\right\}\nonumber
\end{eqnarray}
From this, we may immediately compute correction to the lagrangian and contribution to the entropy using equations (\ref{ES}) and (\ref{1loop}).

\subsection*{Scalar heat kernel for large black holes $qE\gg 1$}

In the following, we describe a different expansion valid in the large black hole regime. Note that in this case, $\rho_s(\lambda)$ behaves like a step function as can be seen from figure~\ref{fig:densities}b (notice the remarkable similarity of this {\it bosonic} spectral density with the density of states for a {\it fermionic} system with chemical potential $\mu=2qE/a$ and temperature $T=1/2\pi a$). Up to exponentially suppressed terms we can approximate
\be
\rho_s(\lambda)=\frac{\sinh(2\pi \lambda)}{\cosh(2\pi \lambda)+\cosh(2\pi qE)}=\frac{1}{1+e^{2\pi(qE-\lambda)}}+{\cal O}(e^{-2\pi qE})
\ee
We can use integration by parts to obtain a useful expression for the $\lambda$-dependent AdS contribution to the heat kernel:
\begin{eqnarray}
\int_0^\infty d\lambda \,\rho_s(\lambda)\,\lambda\, e^{- \bar{s}\lambda^2}&=&\frac{1}{2 \bar{s}}\int_0^\infty d\lambda \,\rho_s'(\lambda)\,e^{- \bar{s}\lambda^2}\approx \frac{\pi}{4\bar{s}}\int_0^\infty d\lambda \,{\text{sech}}^2[\pi(\lambda-qE)]\,e^{- \bar{s}\lambda^2}\nonumber\\
&=& \frac{\pi}{4\bar{s}}\int_{-qE}^\infty dx\,{\text{sech}}^2(\pi\, x)\,e^{- \bar{s}(qE-x)^2}\,,
\end{eqnarray}
where we have defined $x\equiv \lambda - qE$. Because of the $\text{sech}^2(\pi x)$, in any region $|x|\gg 1$ the integrand is exponentially suppressed as $e^{-2\pi x}$. The only contribution comes from $|x|\lesssim 1$, and we can expand the exponent:
\begin{eqnarray}\label{rhosint}
\int_0^\infty d\lambda \,\rho_s(\lambda)\,\lambda\, e^{- \bar{s}\lambda^2}&=& \frac{\pi}{4\bar{s}}\int_{-qE}^\infty dx\,{\text{sech}}^2(\pi\, x)\,e^{- \bar{s}(qE-x)^2}\nonumber\\
&\approx&\frac{\pi}{4\bar{s}} e^{-\bar{s} q^2 E^2} \int_{-qE}^{\infty}dx\, \text{sech}^2(\pi x)\left[1-\bar{s}\left(2qEx+x^2\right)+\frac{\bar{s}^2}{2}\left(2qEx+x^2\right)^2+\ldots\right]\nonumber\\
&=& \frac{\pi}{4\bar{s}}\,e^{-\bar{s} q^2 E^2}\left(\frac{2}{\pi}-\frac{\bar{s}}{6\pi}+\frac{7 \,\bar{s}^2}{240\pi}+\bar{s}^2\frac{q^2E^2}{3\pi}+\ldots\right)
\end{eqnarray}
This expansion is valid in the regime $\bar{s}\ll 1/qE$, or $s\ll a/qM_p$. The most important point to notice from this expression is the exponential suppression factor $e^{-\os q^2 E^2}$. Combining with \eqref{KKtilde}, we see that this precisely cancels the contribution to the effective mass $\Delta m^2$ from the coupling of the particle to the black hole electric field.

Now we see that the scalar effective mass behaves like $\Delta m^{2}$ for $q E \ll 1$, but then transitions to $m^{2}$ for $q E \gg 1$.  This implies that for large black holes the contribution from scalar fields behaves as one would expect in flat space, and no logarithmic corrections arise.  On the other hand, for medium size extremal black holes, extremal particles are effectively massless and may give rise to logarithmic corrections. We will compute these in the remainder of this section.

\subsection*{Fermionic heat kernels}

Let us try to repeat our analysis for fermions. The fermionic heat kernel~\eqref{Kf} may also be divided into medium and large varieties depending on the size of $qE$.  For the medium black holes we may proceed as in (\ref{intermediateKAdS}), but from the starting point:
\be
\rho_{f}(\lambda) = -1 - 2\sum_{k=1}^{\infty} \cosh\left(2\pi k q E\right) e^{-2\pi k \lambda}
\ee
Following an analogous procedure, we are led to the following fermionic heat kernel for intermediate black holes:
\begin{eqnarray}\label{totalkernelf}
K_{\text{medium},f}(s)&=&e^{-s \Delta m^2}\left\{\frac{-1}{8\pi^2 s^2}+\frac{1}{48\pi^2 s}\left(\frac{1}{b^2}-\frac{1}{a^2}+\frac{6 q^2 E^2}{a^2}\right)\right.  \\
&+&\left. \left[\frac{1}{480\pi^2}\left(\frac{1}{a^4}+\frac{1}{b^4}\right)+\frac{1}{288\pi^2 a^2 b^2} -\frac{q^2 E^2}{48\pi^2 a^2 b^2} +\frac{q^2 E^2}{16\pi^2 a^4} - \frac{q^4 E^4}{64\pi^2 a^4}\right] +{\cal O}(s)\right\}\nonumber
\end{eqnarray}

However, one immediately sees a problem once we attempt to probe large black holes and this problem even undermines the validity of the expansion we have written above for medium black holes.  The issue is that the pole in the fermionic spectral density~\eqref{Kf} prevents us from providing a tractable description of the physics in the regime $q E \gg 1$.  At the very least we would need to know something about the UV completion of the theory already at the scale $\lambda = q E$.
 It is reasonable, therefore, to suspect that a consistent UV completion will cut off this theory before $\lambda = q E$ is reached.  Translating this  bound into ordinary energies and assuming that the black hole is well-described by the classical solution, this corresponds to an energy of:
\be\label{WGCcutoff}
\text{Energy} \sim \frac{\lambda}{a} \sim \sqrt{2} q M_{P}
\ee
Note that this is precisely the threshold for the magnetic Weak Gravity Conjecture argued for via other means in \cite{ArkaniHamed:2006dz}.  

While we find this to be a suggestive argument for the magnetic WGC cutoff,  one might nevertheless object by saying that this could equally well be interpreted as an IR cutoff. This should be implemented by setting a lower bound in the $\lambda$-ingetration in~\eqref{1loop} to $\lambda\gtrsim qE$ or $\lambda/a\lesssim qM_P$.  This prescription would allow us to evade the ambiguity from the fermionic heat kernel pole. However, we cannot find a good fundamental motivation for such an IR cutoff. In fact, this cutoff would remove the dominant low-momentum contributions to one-loop amplitudes, e.g. in flat space, and standard results from renormalization theory might be affected. We find more appealing the interpretation of~\eqref{WGCcutoff} as a UV cutoff, in agreement with the WGC (c.f. section~\ref{comments}).   Of course, more work will be needed to fully understand the significance of this pole. 

While the pole in the fermionic spectral density poses an impasse for the study of large black holes, we may still study the corrections to the entropy of intermediate black holes for which the approximations~\eqref{totalkernels} and~\eqref{totalkernelf} can be used. In the following, we will study the corrections induced by subextremal and extremal particles separately.

\subsection{Subextremal Particles: $m^2>2q^2M_P^2$}
Having now developed the heat kernels in the previous sections, we are ready to write down the corrections to the entropy using equations (\ref{ES}) and (\ref{1loop}).  For intermediate black holes with sub-extremal particles one finds the 1-loop corrections:

\begin{eqnarray}\label{chargedsubL1s}
{\mathcal L}_s^{(1)}&=&  \frac{1}{32\pi^2 \epsilon^4} -\frac{m^2}{16\pi^2\epsilon^2}+\frac{1}{48\pi^2 \epsilon^2}\left(\frac{1}{b^2}-\frac{1}{a^2}\right) \\
&& -\left[\frac{m^4}{32\pi^2}  -\frac{m^2}{48 \pi^2}\left(\frac{1}{b^2}-\frac{1}{a^2}\right) +\frac{1}{240\pi^2}\left(\frac{1}{a^4}+\frac{1}{b^4}\right)-\frac{1}{144\pi^2 a^2 b^2}+\frac{q^2 E^2}{96 \pi^2 a^4} \right]\ln{\left(\epsilon^2\Delta m^2\right)}\nonumber
\end{eqnarray}

\begin{eqnarray}\label{chargedsubL1f}
{\mathcal L}_f^{(1)}&=&  -\frac{1}{32\pi^2 \epsilon^4} +\frac{m^2}{16\pi^2\epsilon^2}+\frac{1}{96\pi^2 \epsilon^2}\left(\frac{1}{b^2}-\frac{1}{a^2}\right) \\
&& +\left[\frac{m^4}{32\pi^2}  +\frac{m^2}{96 \pi^2}\left(\frac{1}{b^2}-\frac{1}{a^2}\right) -\frac{1}{960\pi^2}\left(\frac{1}{a^4}+\frac{1}{b^4}\right)-\frac{1}{576\pi^2 a^2 b^2}-\frac{q^2 E^2}{48 \pi^2 a^4} \right]\ln{\left(\epsilon^2\Delta m^2\right)}\nonumber
\end{eqnarray}
Combining this one loop contribution with the classical term~\eqref{lag}, one can check the proper cancellation of $UV$ divergent terms against counterterms, and obtain finite entropy functions.

We can estimate the quantum corrected black hole entropy by evaluating the entropy function at the classical solution~\eqref{tree}. In this way, we find the quantum corrected entropies:
\begin{eqnarray}
{\mathcal S}_{s}&\approx&{\mathcal E}(Q;E_0,a_0,b_0)=\frac{Q^2}{4}+\left(\frac{m^4Q^4}{(8\pi)^4M_P^4}+\frac{q^2Q^2}{192\pi^2}\right)\ln\left(1-\frac{2q^2M_P^2}{m^2}\right)\nonumber\\
&\approx& Q^2\left[\frac{1}{4}-q^2 Q^2 \frac{m^2}{(8\pi)^4 M_p^2}\right]+{\cal{O}}\left(q^4Q^2, \frac{q^{2}M_{P}^{2}}{m^{2}}\right)
\end{eqnarray}
\begin{eqnarray}
{\mathcal S}_{f}&\approx&{\mathcal E}(Q;E_0,a_0,b_0)=\frac{Q^2}{4}+\left(-\frac{m^4Q^4}{(8\pi)^4M_P^4}+\frac{q^2Q^2}{96\pi^2}\right)\ln\left(1-\frac{2q^2M_P^2}{m^2}\right)\nonumber\\
&\approx& Q^2\left[\frac{1}{4}+q^2 Q^2 \frac{m^2}{(8\pi)^4 M_p^2}\right]+{\cal{O}}\left(q^4Q^2,\frac{q^{2}M_{P}^{2}}{m^{2}}\right)
\end{eqnarray}
The reader should recall that this result is only valid for intermediate-size black holes for which $1\ll Q\ll 1/q $. Hence, the contributions from masive sub-extremal particles, although formally proportional to $Q^4$, are small corrections to the Beckenstein-Hawking term.

Notice that the  terms scaling as $\sim Q^4$ arise from the correction to the lagrangian~\eqref{chargedsubL1s} and~\eqref{chargedsubL1f} proportional to $m^4 \log(\Delta m^2 \epsilon^2)$. The logarithmic divergence in these terms corresponds to the renormalization of the cosmological constant\footnote{For details of the renormalization see eq.~\eqref{renormalized}}.   The remaining, finite, piece depends on the fields and is therefore physical.  It may be thought of as analogous to a cosmological constant term in the following sense.  From equations (\ref{Kb}) and (\ref{Kf}) one can see that by turning on charge, we have suppressed the density of states for small $\lambda$ by $e^{-2\pi q E}$.  Here, `small' means $\lambda < q E$, which corresponds to physical energies $\lesssim q M_{P}$. Thus, the charge has effectively projected out all the sub-planckian modes, giving rise to an `un-renormalization'\footnote{i.e., the usual flat space renormalization of the cosmological constant is precisely accounted for when all these modes are not suppressed, so suppressing them is `un-renormalizing', thus revealing the large bare term.} of the cosmological constant.

One may expect, then, that in a theory in which the cosmological constant is not renormalized these corrections to the entropy will not arise. In fact, if we consider the contribution to the black hole entropy from a chiral ${\cal N}=1$ multiplet, by simply adding the corrections from \eqref{chargedsubL1s} and~\eqref{chargedsubL1f}, we find a black hole entropy of
\be\label{SN=1}
{\mathcal S}_{{\cal N}=1}\approx{\mathcal E}(Q;E_0,a_0,b_0)=\frac{Q^2}{4}+\frac{q^2Q^2}{64\pi^2}\ln\left(1-\frac{2q^2M_P^2}{m^2}\right)+\mathcal{O}(Q^{0})
\ee
The $Q^4$ term has indeed disappeared. The leading correction, although entering at order $Q^2$, is suppressed with respect to the Beckenstein-Hawking term. In summary, we may conclude that 
the entropy in the supersymmetric case is the Beckenstein-Hawking entropy, plus small corrections:
\be
\mathcal{S}_{{\cal N}=1} = \frac{Q^{2}}{4}\left(1+\mathcal{O}\left(\frac{q^{4} M_{P}^{2}}{m^{2}}\right)\right)
\ee
As we approach the extremal bound the perturbative expansion in $q$ ceases to converge and we must use an alternative approach.  This will be discussed next.

\subsection{Extremal Particles: $m^2=2q^2 M_P^2$}
The case with extremal particles is particularly important, 
especially when one has coincident BPS and extremality bounds.  Such situations commonly arise in KK reductions of supersymmetric theories where the internal KK momentum serves as a central charge with respect to the KK $U(1)$.  In this case the particles behave effectively as massless particles in the regime of intermediate black holes and can lead to corrections which contain a logarithmic scaling.  We note that the log term is a robust prediction, while higher order terms are subject to UV corrections.  In order to extract the log, we proceed as in previous sections; i.e., extract the leading terms in the small $s$ expansion of the heat kernel and then use (\ref{1loop}).  The suppresion at large $s$ now comes from the factors like $e^{-s/4a^{2}}$.  

The 1-loop corrections for extremal particles in the background of intermediate black holes may be obtained simply by replacing $\ln(\epsilon^{2} \Delta m^2) \rightarrow \ln(\epsilon^{2}/a^{2})$ and $m\rightarrow \sqrt{2} q M_P$ in the corresponding equations for sub-extremal particles.  After performing this substitution and proceeding as before one finds:
\begin{eqnarray}\label{entrexts}
{\mathcal E}_s&=& 2 \pi\left\{Q E - 4\pi a^2\, b^2\,\left[
 M_P^2\left(\frac{1}{b^2}-\frac{1}{a^2}\right)+\frac{E^2}{2a^4}\right.\right.\\
&&+\left.\left.\left(\frac{m^4}{32\pi^2}  -\frac{m^2}{48 \pi^2}\left(\frac{1}{b^2}-\frac{1}{a^2}\right) +\frac{1}{240\pi^2}\left(\frac{1}{a^4}+\frac{1}{b^4}\right)-\frac{1}{144\pi^2 a^2 b^2}+\frac{q^2 E^2}{96 \pi^2 a^4} \right)\ln{(a^2 m^2)}\right]\right\}\nonumber
\end{eqnarray}
\begin{eqnarray}\label{entrextf}
{\mathcal E}_f&=& 2 \pi\left\{Q E - 4\pi a^2\, b^2\,\left[
 M_P^2\left(\frac{1}{b^2}-\frac{1}{a^2}\right)+\frac{E^2}{2a^4}\right.\right.\\
&&-\left.\left.\left(\frac{m^4}{32\pi^2}  +\frac{m^2}{96 \pi^2}\left(\frac{1}{b^2}-\frac{1}{a^2}\right) -\frac{1}{960\pi^2}\left(\frac{1}{a^4}+\frac{1}{b^4}\right)-\frac{1}{576\pi^2 a^2 b^2}-\frac{q^2 E^2}{48 \pi^2 a^4} \right)\ln{(b^2 m^2)}\right]\right\}\nonumber
\end{eqnarray}
The first line in each of these expressions represents the classical entropy function, written in terms  of the renormalized Planck mass and gauge field strength. The $\log$ terms in the second lines, although reminiscent of renormalization, are actually coming from a resummation of the corrections appearing in $\ln(\Delta m/m)$ and should therefore be thought of as a physical correction.

The quantum entropy function evaluated at the classical solution is:
\be
{\mathcal S}_{s}\approx {\mathcal E}(Q;E_0,a_0,b_0)=\frac{Q^2}{4}-\left(\frac{1}{90}+\frac{q^2Q^2}{192\pi^2}+\frac{q^4Q^4}{1024\pi^4}\right)\ln(q^{2}Q^2)+\mathcal{O}(Q^{0})
\ee
with a similar result for fermions.  The corrections are small for $q Q \ll 1$, which is the only regime in which this result is to be trusted.  As $q Q \rightarrow 1$, the exponential suppression factor transitions from $e^{-s /4 a^{2}}$ to $e^{-s m^{2}}$, as demonstrated in \eqref{rhosint}

We may also write down the result for a minimally coupled $\mathcal{N}=1$ supersymmetric particle in this background simply by adding the contributions from a (complex) scalar and a charged (Weyl) fermion: $K_{{\cal N}=1}=K_s+K_f$.  For $qQ \ll 1$, one finds:
\be\label{susyextrresult}
{\mathcal S}_{{\cal N}=1}\approx{\mathcal E}(Q;E_0,a_0,b_0)=\frac{Q^2}{4}-\left(\frac{q^2Q^2}{64\pi^2}+\frac{1}{24}\right)\ln(q^{2}Q^2)+\mathcal{O}(Q^{0})
\ee
\subsection{Comments On Cutoffs}\label{comments}
So far, we have been operating under the assumption that the UV cutoff, $\epsilon$, is independent of $q$ and $M_{P}$ and may be taken to zero.  In this case, we can unambiguously identify divergences and cancel them with counterterms.  However, the pole in the fermionic density of states, as well as the arguments of \cite{ArkaniHamed:2006dz}, suggest that we should in fact place our cutoff at or below a UV scale of $q M_{P}$.  However, if the cutoff were to be placed below $q M_{P}$, it would also be below the mass scale of the particles we are integrating out.  In this case, we should not integrate out these particles at all, rather, if such particles are present, we should think of them as solitonic objects in the low energy effective field theory.

\section{Generalizations}
\label{general}

\subsection{Dyonic Black Holes}

In this section, we describe generalizations of the one-loop calculations to dyonic black holes and theories with multiple $U(1)$'s as this may be important for more realistic stringy realizations such as those considered in \cite{Sen:2007qy}.  In fact, when a massless dilaton is present, the dyonic property is often necessary in order to guarantee that the black hole area is non-zero and this fact is likely to be preserved under generic perturbations of the theory.    We therefore wish to consider the effect of adding a dyonic particle to the background of a dyonic black hole.\footnote{Here we will only consider the simplest case of Einstein-Maxwell with a single $U(1)$, but the generalization to multiple $U(1)$'s with generic kinetic mixing is straightforward.}  We begin with the classical background (\ref{bg}) described as follows:

\be
F \rightarrow E dt \wedge dr + B \sin\theta d\theta \wedge d\phi
\ee
This modifies the classical lagrangian density and solution to:  
\beq
&\mathcal{L}^{(0)}_{GR+EM} &\sim \frac{E^{2}}{2 a^{4}} - \frac{B^{2}}{2 b^{4}} +M_{P}^{2} \left(\frac{1}{b^2}-\frac{1}{a^2}\right) \\ \nonumber
&\Longrightarrow & 
a_0^2 = b_0^2 = \frac{E^{2}+B^{2}}{2 M^{2}_{P}}  \\ \nonumber
\eeq
We find an extremal mass of:
\be
\label{mdyon}
M_{extremal} = 4\pi\sqrt{2}  M_{P} \sqrt{E^{2}+B^{2}}
\ee

If the fundamental unit of charge is $q$,  Dirac quantization imposes the following conditions on $E$ and $B$:
\be
E = \frac{q N_{e}}{4\pi}\,\,\,\,\,\,\,\,\,\,\,\,\,\,\, B = \frac{N_{m}}{2 q}
\ee
where $N_{e},\, N_{m} \in \mathbb{Z}$ and $q N_{e} = Q$ in terms of our earlier solution.

Consider now adding a particle with $n_{e}$ units of electric charge and $n_{m}$ units of magnetic charge.  The net effect of the charge interaction is to shift the angular momentum via \cite{hald} $l \rightarrow l+s_{0}$ where $s_{0} \equiv \frac{1}{2}|n_{e} N_{m}- n_{m}N_{e}|$.  In addition the spherical eigenstates are modified via:
\be
\mathbf{L}^{2} = \frac{l(l+1)}{b^{2}} \rightarrow \frac{1}{b^{2}}\left(l(l+1) +2 s_{0}\left(l+\frac{1}{2}\right)\right)
\ee
We thus expect the (complex scalar) heat kernel to be modified to:
\beq
K^{s}_{q}(s) &=&\frac{e^{-s \Delta m^{2}}}{4\pi^{2}a^{2} b^{2} }\sum_{l=0}^{\infty}(2l+2 s_{0} +1)\int_{0}^{\infty} d\lambda\, \lambda\,\rho(\lambda)e^{-s\left(\left(\lambda^{2}+\frac{1}{4}\right)/a^2+\left(l(l+1)+2s_{0}\left(l+\frac{1}{2}\right)\right)/b^2\right)} \nonumber \\
\rho(\lambda) &=& \frac{\sinh(2\pi \lambda)}{\cosh (2\pi \lambda)+\cosh\left(2\pi |n_{e} q E+2\pi n_{m}B/q|\right)} 
\eeq
where
\be
\Delta m^{2} = m^{2} - \frac{\left(n_{e} q E + \frac{2\pi n_{m}B}{q}\right)^{2}}{a^{2}}
\ee
Similar results hold for the fermions.  From the above formulas it is easy to check that:
\be
a^{2} \Delta m^{2} = s_{0}
\ee
Notice that $s_{0}$ is zero precisely when the charges of the black hole and particle are aligned, which  classically reduces to the case of pure electric charge.  Moreover, note that $s_{0}$ has the interpretation of minimal angular momentum of the system.  It is not surprising that this extra energy would drive the dyonic field to be effectively subextremal.

\subsection{Multi-centered black holes and primitive charges}
The extremal black holes we have considered so far are energetically allowed to split into smaller black holes.\footnote{The phase space for such a dynamical decay starting from an extremal state is nevertheless empty.  Furthermore, if the entropy of an extremal black hole of charge $Q$ grows as $S\sim Q^2$ (the classical result), the black hole is  entropically bound. That is, its decay into two black holes of charges $Q_1$ and $Q_2$ (with $Q=Q_1+Q_2$) is entropically disfavored, since $S_Q> S_{Q_1}+S_{Q_2}$.} This raises the question of whether the calculations presented in this note are robust to fractionation. One way to address this issue is to consider extremal black holes charged under two $U(1)$s, with primitive charge vector $\vec{Q}=(Q_1,Q_2)$, with $Q_1$ and $Q_2$ mutually prime charges. If the mass of the black hole grows as $M^2\sim Q_1^2+Q_2^2$ (the classical result in Einstein-Maxwell theory), these primitively charged black holes are safely bound\footnote{In setups with a massless dilaton, a different behavior of the mass can arise $M\sim |Q_1|+|Q_2|$. In this case, not even primitively charged extremal black holes are safely bound.}.

We can hence consider the problem of integrating out a field of charge $\vec{q}=(q,0)$ in the near horizon geometry of an stable extremal black hole of charge $\vec{Q}=(Q_1,Q_2)$ that generates an electric field $\vec{E}\propto \vec{Q}$. We can study the case $Q_2\ll Q_1$. At the classical level, extremality of the black hole implies
\begin{equation}
E_1^2+E_2^2=2M_p^2 a^2\qquad \Longrightarrow\qquad E_1^2=2M_p^2 a^2 -E_2^2
\end{equation}

All the previous calculations go through identically with the replacement $q E\to q E_1$. In particular, the exponential suppression of the heat kernel is given by 
\begin{equation}
\Delta m^2=m^2-\frac{q^2 E_1^2}{a^2}\,,
\end{equation}
which still has the same scaling behavior as in the singly-charged black hole case
\begin{equation}
\Delta m^2=m^2-2q^2  M_p^2+\frac{q^2 E_2^2}{a^2}=m^2- 2 q^2 M_p^2+{\cal O}(a^{-2})\,.
\end{equation}
In particular, an extremal particle will still behave as effectively massless and induce corrections proportional to $\log a$ while we have avoided the issue of fractionation.

\subsection{String theory black holes}
\label{sbh}
It is natural to ask how the calculations presented here relate to black holes appearing in string theory.  Given the succesful matching between leading order corrections to the Beckenstein-Hawking entropy and microscopic descriptions in certain theories \cite{Strominger:1996sh, Sen:2007qy, Dabholkar:2012zz}, we will assume the classical description to be a good approximation in these cases. It is, of course, possible that the classical description breaks down at the horizon as is advocated by many authors\footnote{See, for instance \cite{Mathur:2005zp, Almheiri:2012rt}.}.  However, since we are interested in probing the `intrinsic' limits of effective field theory, we would like to understand what ingredients make it possible for EFT to be succesful despite possible tensions with fuzzballs/firewalls\footnote{Another possible way of phrasing the question is `for which EFTs can fuzzball complentarity work?'.}.  In this context, we note that extremal particles are fairly common in string theories compactified down to 4 or 5 dimensions, and we have seen that loops of these can induce important modifications to the classical description of black holes. It would thus be interesting to explore how these considerations affect string theoretic results.  

The stringy black holes with which we make the closest comparison are the black holes described by \cite{Sen:2007qy}, which have both an $AdS_{2}\times S^{2}$ geometry and pointlike elementary degrees of freedom\footnote{Other classes of stringy black holes, for instance, those of \cite{Benini:2016rke} have a near horizon description that is $AdS_{2}\times S^{2}$, but the microscopic degrees of freedom are described by an $M2$ brane.}.    These are dyonic black holes which may arise from a number of different string constructions, have either $\mathcal{N}=4$ or $\mathcal{N}=2$ susy in 4d and generically admit a number of $U(1)$ gauge fields at generic points in the moduli space.  At special points in the moduli space the gauge group is enhanced to a non-abelian group and there are consequently massive BPS `W-boson' multiplets associated with Higgsing.  In principle, one might worry that loops of such particles could give rise to large corrections.  

Notice, first of all, that such theories contain BPS bounds, that strictly forbid the existence of super-extremal particles. Black holes are hence not allowed to emit their charge away via Schwinger pair production. On the other hand, these string compactifications also contain a plethora of new ingredients in their low energy effective action (in particular massless axio-dilatons). These extra elements not only enter the black hole solutions, but also modify the kinetic operators of matter fields, and hence their heat kernel. Furthermore, the smallest charged multiplet that one can consider in an $\mathcal{N}=4$ theory contains not only scalars and fermions, but also vector bosons. In order to compute the quantum corrections to extremal black holes in this case, we would need to generalize the computations of the previous sections to include such fields, and in particular, compute the heat kernel of a (massive) vector boson. 

However, at least for $\mathcal{N}=4$ theories, the high degree of supersymmetry plays an important role in keeping loop corrections under control.  For instance, in \cite{Banerjee:2010qc} it was shown that loops of massless vector multiplets in $\mathcal{N}=4$ supergravity give zero contribution to the effective action.  Moreover, it has been argued\footnote{See, e.g., \cite{Kallosh:1992ii, Duff:1982yw, Gates:1983nr}.} on general grounds that in $\mathcal{N}=4$ supergravities the classical action does not renormalize\footnote{Note that Strominger-Vafa black holes are $\mathcal{N}=4$ in 5d with a near horizon geometry of $AdS_{2}\times S^{3}$ and so are likewise protected \cite{Sen:2012cj}.}. Notice that this is due to particular couplings of such fields to the gauge fluxes and geometry of the black hole. For example, a ten dimensional coupling of the form $\bar{\Psi} \Gamma^{MNP}\Psi H_{MNP}$ between fermions and the three-form field strength leads to a modification of the four dimensional fermionic kinetic operator roughly of the form $\gamma^\mu D_\mu \to \gamma^\mu D_\mu +\gamma^{\mu\nu}F_{\mu\nu}$. This, and similar contributions, conspire to make the total heat kernel vanish, $K_{{\cal N}=4}=0$, something that does not necessarily happen in simpler theories.

It is interesting to consider the case of ${\cal N}=2$ supergravity, whose charged chiral multiplets consist solely of scalars and fermions. Just as with $\mathcal{N}=4$ supersymmetry, there exists a BPS bound that forbids the existence of super-extremal states. All fields must be either extremal or sub-extremal.   As we have seen in the previous section, minimally coupled extremal fields can induce logarithmic corrections to the black hole entropy.   Using the techniques of~\cite{Keeler:2014bra,Larsen:2014bqa}, one can in fact show that similar corrections are still present \cite{wip} when non-minimal coupling is considered.   One might therefore suspect that we are not supposed to integrate out these extremal particles at all.  In fact, this is a lesson one may draw from the resolution of the conifold singularity \cite{Strominger:1995cz}.  In this case, one sees that the low energy stringy effective action behaves as if the extremal D-brane states had {\it{already}} been integrated out, thus giving rise to a conifold singularity.  However, the task of clarifying whether or not this stands as a universal prescription, or, more generally which ingredients play a role in the cancellation of the corrections still remains. We leave such studies for future work.

\section{Speculations / Conclusion}
An analysis of systems of subextremal particles in either the classical or quantum mechanical regime teaches us that the microcanonical entropy of bound states is divergent for sufficiently large systems of particles.  However, looking at the classical general relativistic description, we are unable to detect this entropy due to the presence of a horizon.  This raises the question as to what has happened to the classically expected entropy.  Is the horizon cloaking a much larger entropy, and, if so, could such a theory be consistent with unitarity?  

We have attempted to study this problem by looking at one-loop corrections to the near-horizon geometry.   While our results do not establish that sub-extremal particles contribute a divergent entanglement entropy, this does not directly say anything about the microcanonical entropy of the system.

\begin{figure}[h]\center{
\includegraphics[scale = .4]{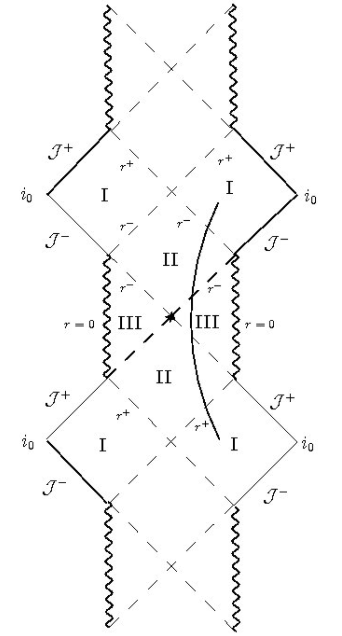}}
\caption{Maximal extension of the Reissner-Nordstrom metric reproduced from \cite{hawking1975large}.  The diagram repeats an infinite number of times in the vertical direction.}
\label{RNpenrose}
\end{figure}

Indeed, there is reason to believe that the entanglement entropy calculated here is simply insufficient to represent the full microcanonical entropy.  If one looks, for instance, at the extended Penrose diagram of the Reissner-Nordstrom black hole (Fig.~\ref{RNpenrose}), one sees that the horizon does not form a Cauchy surface for the interior region and therefore does not contain all the relevant information.  There is an additional infinite series of time-like horizons with boundary conditions which must be specified.  

A related problem is that in such a spacetime, an observer could in principle travel from one asymptotic region I, through regions II and III all the way to the second region I.  In the process, he would witness the timelike naked singularity at $r=0$; a violation of strong cosmic censorship. This is just another manifestation of the fact that not all of the information has been accounted for on the Cauchy surface. 

For such a Reissner-Nordstrom black hole, if we define the entropy as the number of internal states consistent with external data then the result would appear to be infinite; an arbitrary amount of information must be specified in addition to the horizon data.  This would then be in agreement with our microscopic expectations for a system of subextremal particles. 

One may object that the Penrose diagram is not valid anyway due to the instability of the inner horizon.  However, given that the extended Penrose diagram is a solution, we may use it as a starting point for counting perturbations. There are already an infinite number of perturbations that leave the nearest inner horizon fixed; just consider modes in a distant asymptotic region.  Thus, it is not clear how or if the instability will affect the argument that the entropy diverges.  Furthermore, we note that the inner horizon merely collapses to a weak null singularity - still violating the strong cosmic censorship conjecture \cite{daf1,Dafermos:2004jp, Dafermos2005}. Such a singularity may in fact be considered traversable since the integrated tidal forces are finite.  The relationship between cosmic censorship and WGC has long been suspected \cite{madrid} and recently made more precise in \cite{Horowitz:2016ezu}.  

What the above reasoning suggest is that a hypothetical theory built from subextremal particles will, in fact, have divergent microcanonical entropies though this is hidden inside the horizon and undetectable as entanglement entropy.  Instead, the problem emerges for observers who jump inside the black hole and observe the additional entropy.  This additional entropy is represented as a naked singularity in the Penrose diagram.  More generally, the entropy is associated with the non-uniqueness of the maximal extension of the Cauchy data associated with the outer horizon.  However, it would be very difficult to formalize this claim without a good measure on this space of solutions in general relativity.  

It is difficult to imagine a universe where semi-classical Reissner-Nordstrom black holes having the properties described above exist.  For starters, observers in any given asymptotic region would not have a unitary description of events since information could effectively leave their universe and wind up in another asymptotic region.  There is no way for the information to come back as Hawking radiation, nor as Schwinger particles if WGC is violated.   Secondly, the usual anti-remnant argument \cite{Susskind:1995da} applies if the entropy of classical extremal black holes diverges, as the arguments presented here suggest.  Even if the microcanonical entropy were to increase by some power faster than the area, rather than precisely as the area, then loops of large black holes would dominate the Rindler vacuum at small acceleration following the same reasoning.  This would imply that low temperature physics is dominated by black holes.

The most elegant way to resolve these problems is just to postulate the existence of at least one particle satisfying $\Delta m^{2} \le 0$ as originally suggested \cite{ArkaniHamed:2006dz}.  In this work, we have attempted to provide further arguments that these issues indeed strongly merit a resolution.  

To further diagnose the `sickness' associated with theories violating the WGC, it would be useful to have a better characterization of the statistics of bound states of mutually attractive particles {\textit{after}} the formation of a black hole.  This would necessarily entail going beyond the realm of equilibrium thermodynamics.  It would also be interesting to explore the information theoretic content of the inner horizons and their relationship to the entropy as measured by external observers.  We leave this for future work.

\paragraph{Acknowledgements}
We thank  Jos\'e Luis F. Barb\'{o}n, Jon Brown, Anthony Charles, Arthur Hebecker, Aitor Landete, Finn Larsen, Miguel Montero, Eran Palti, Angel Uranga and especially Ashoke Sen for useful discussions and comments. This work was supported in part by the DOE grant DE-FG-02-95ER40896, the Kellett Award of the University of Wisconsin, and the HKRGC grants HUKST4/CRF/13G, 604231 and 16304414. Support for W.C.'s research was also provided by the Graduate School and the Office of the Vice Chancellor for Research and Graduate Education at the University of Wisconsin-Madison with funding from the Wisconsin Alumni Research Foundation. P.S. acknowledges support from the DFG Transregional Collaborative Research Centre TRR 33 ``The Dark Universe".

\appendix
\section{Counterterms and renormalization of effective action in flat space}

In this appendix we summarize one-loop renormalization results for Einstein-Maxwell theory in flat spacetime. 
Let us start with Einstein-Maxwell theory, with an Euclidean action given by
\begin{eqnarray}
S^{(0)}&=&\frac{1}{16\pi G_N^{(0)}}\int d^4x\sqrt{g}\left({\cal R}-2\lambda^{(0)}\right)\\
&+&\int d^4x\sqrt{g}\left(c^{(0)}_1 {\cal R}_{;\mu}^{\,\,\,\,\mu}+c^{(0)}_2{\cal R}^2+c^{(0)}_3{\cal R}_{\mu\nu}{\cal R}^{\mu\nu}+c^{(0)}_4{\cal R}_{\mu\nu\rho\sigma}{\cal R}^{\mu\nu\rho\sigma}-\frac{1}{4}F^{(0)}_{\mu\nu}F^{(0)\,\mu\nu}\right)\nonumber
\end{eqnarray}
where the $(0)$ superscript indicates {\it bare} parameters. The one-loop contribution from a scalar of mass $m$ is given by:
\begin{equation}
W^{(1)}_{eff}=\frac{1}{2}\int_{\epsilon^2}^\infty \frac{ds}{s}\,  \tilde{K}(s) \,e^{-s\,m^2}\,,
\end{equation}
where  $\tilde{K}(s)$ is the trace of the heat kernel of the kinetic operator in flat space, and $\epsilon$ is the short distance cutoff. The leading terms of the heat kernel, for a scalar field of charge $q$, can be read, e.g., from the general formulae in~\cite{Vassilevich:2003xt}:
\begin{eqnarray}
\tilde{K}_s(s)&=& \frac{1}{8\pi^2 s^2}\int d^4x\sqrt{g} + \frac{1}{48\pi^2 s}\int d^4x \sqrt{g}{\cal R}\nonumber\\
&+& \frac{1}{2880\pi^2}\int d^4 x \sqrt{g}\left(-12 {\cal R}_{;\mu}^{\,\,\,\,\mu}+5{\cal R}^2-2{\cal R}_{\mu\nu}{\cal R}^{\mu\nu}+2{\cal R}_{\mu\nu\rho\sigma}{\cal R}^{\mu\nu\rho\sigma}-30 q^{2} F_{\mu\nu}F^{\mu\nu}\right)\nonumber
\end{eqnarray}
From these, we can obtain the standard results for the renormalized couplings:
\begin{eqnarray}\label{renormalized}
\frac{1}{G_N^{(r)}}&=&\frac{1}{G_N^{(0)}}+\frac{1}{6\pi\epsilon^2}+\frac{m^2}{6\pi}\ln{(\epsilon m)} +{\cal O}(\epsilon^0)\nonumber\\
\frac{\lambda^{(r)}}{G_N^{(r)}}&=&\frac{\lambda^{(0)}}{G_N^{(0)}}-\frac{1}{4\pi\epsilon^4}+\frac{m^2}{2\pi\epsilon^2}+\frac{m^4}{4 \pi}\ln{(\epsilon m)} +{\cal O}(\epsilon^0) \nonumber\\
c_1^{(r)}&=&c_1^{(0)}+\frac{1}{480\pi^2}\ln{(\epsilon m )} +{\cal O}(\epsilon^0)\nonumber\\
c_2^{(r)}&=&c_2^{(0)}-\frac{1}{1152\pi^2}\ln{(\epsilon m)} +{\cal O}(\epsilon^0)\nonumber\\
c_3^{(r)}&=&c_3^{(0)}+\frac{1}{2880\pi^2}\ln{(\epsilon m)} +{\cal O}(\epsilon^0)\nonumber\\
c_4^{(r)}&=&c_4^{(0)}-\frac{1}{2880\pi^2}\ln{(\epsilon m)} +{\cal O}(\epsilon^0)\nonumber\\
A_\mu^{(r)}&=&A_\mu^{(0)}\left(1-\frac{q^2}{48 \pi^{2}}\ln{(\epsilon m)}\right)^{1/2}+{\cal O}(\epsilon^0)
\end{eqnarray}
In the massless case $m=0$ in which there is no $e^{-sm^2}$ suppression in the heat kernel, one must introduce an arbitrary scale $\mu$. The renormalized quantities take the same form as above, except that the logarithms must be replaced $\ln(\epsilon m)\to\ln(\epsilon\mu)$.  

One may now compare these renormalized quantities with those read off from (\ref{chargedsubL1s}).  The results are in agreement, thus providing a non-trivial check on our calculations, and in particular of the form of the heat kernel~\eqref{Kb}  of the near horizon geometry.  Likewise, we may also deduce the correction coming from a chiral fermion using the heat kernel below:
\begin{eqnarray}
\tilde{K}_f(s)&=& -\frac{1}{8\pi^2 s^2}\int d^4x\sqrt{g} + \frac{1}{96\pi^2 s}\int d^4x \sqrt{g}{\cal R}\nonumber\\
&+& \frac{1}{2880\pi^2}\int d^4 x \sqrt{g}\left(3 {\cal R}_{;\mu}^{\,\,\,\,\mu}+\frac{2}{3}{\cal R}^2-2{\cal R}_{\mu\nu}{\cal R}^{\mu\nu}+\frac{1}{2}{\cal R}_{\mu\nu\rho\sigma}{\cal R}^{\mu\nu\rho\sigma}-60 q^{2} F_{\mu\nu}F^{\mu\nu}\right)\nonumber
\end{eqnarray}
Again, we find agreement between the known renormalization and that which is inferred from (\ref{chargedsubL1f}).


\end{document}